LBNL Physics Note                                                                                        12.26.01

**Semi-classical derivation of the number of photons emitted by a neutrino with mass and magnetic moment passing through a magnetic field**

J. D. Jackson

## A. Preliminaries

We consider a neutrino of mass $m$, energy $E = \gamma mc^2$, and magnetic moment of magnitude $\mu$ moving at uniform speed $v \approx c$ in the $z$ direction in the laboratory in which there is a magnetic field in the $x$ direction of magnitude $B(z)$. For conceptual purposes we imagine a uniform field $B_0$ of length $L$ in the $z$ direction, but all that enters to a high degree of accuracy is

$$\int |B(z)|^2 dz = B_0^2 L \tag{1}$$

The parameters apparently of interest are typically $m = 1$ eV, $E = 50$ GeV, $L = 10$ m, and $B_0 = 2$ tesla. [Ref. 1]. The neutrino is therefore *extremely* relativistic ($\gamma = O(5 \cdot 10^{10})$).

---

[1] K. S. McFarland, A. C. Melissinos, N. V. Mikheev, and W. W. Repko, "Neutrino scattering in a magnetic field," (Rochester preprint, September 25, 2001)

---

We perform our calculation in the neutrino's rest frame K'. In that frame, the transverse magnetic field becomes $B'(z') = \gamma B(z'=z/\gamma)$, accompanied by a transverse electric field in the $y$ direction $E' \approx B'$. The neutrino at rest 'sees' a pulse of radiation traveling in the negative $z$ direction, with time-varying fields, $B_x'(t' = z/\gamma c)$, $E_y'(t' = z/\gamma c)$. The frequency spectrum of the pulse of virtual quanta is given in terms of

$$B'(\omega') = \frac{1}{\sqrt{2\pi}} \int dt' B'_x(t') e^{i\omega' t'} \tag{2}$$

We have the identity,

$$\int d\omega' |B'(\omega')|^2 = \int dt' |B'(t')|^2 = \gamma B_0^2 L / c \tag{3}$$

Here we have one power of $\gamma$ relative to (1) because of the Lorentz transformation properties of $B_x$ and of the coordinate $z$. Below we need the one-sided frequency integral,

$$\int_0^\infty d\omega' |B'(\omega')|^2 = \frac{1}{2} \gamma \ B_0^2 L/c \tag{4}$$

With the typical parameters given above, the range of support in the incident virtual photon's energy is of order $\hbar \gamma c / L = O(10^3 \text{ eV})$. The typical frequency involved in the scattering or induced radiation by the neutrino therefore has $\hbar \omega'_{typical} >> mc^2$. Whatever the shape of the magnetic field in the laboratory, the pulse of virtual photons will appear as almost a delta function in rest-frame time to the neutrino.

**B. Classical intensity of radiation**

Classically, the power radiated per unit solid angle in K' by the driven magnetic moment of the neutrino is [ref. 2]

---

[2] J. D. Jackson, *Classical Electrodynamics*, 3rd ed., (Wiley, N.Y., 1998), p. 451, problem 9.7(a).

---

$$\frac{dP(t')}{d\Omega'} = \frac{1}{4\pi c^3} \left| \hat{\mathbf{k}} \times \frac{d^2 \mathbf{m}}{dt'^2} \right|^2 \tag{5}$$

Here $\hat{\mathbf{k}}$ is a unit vector in the direction of the emitted radiation, **m** is the neutrino's magnetic moment, and contrary to ref. 2 the power is expressed in Gaussian units.

The equation of motion for the magnetic moment in an external field is found from the torque equation for its spin *s*, using the relation $\mathbf{m} = 2\mu \mathbf{s}/\hbar$:

$$\frac{d\mathbf{m}}{dt'} = \frac{2\mu}{\hbar} \ \mathbf{m} \times \mathbf{B}'(t') \tag{6}$$

The second time derivative of **m** is

$$\frac{d^2 \mathbf{m}}{dt'^2} = \frac{2\mu}{\hbar} \ \mathbf{m} \times \frac{d\mathbf{B}'(t')}{dt'} + \frac{d\mathbf{m}}{dt'} \times \mathbf{B}'(t') \tag{7}$$

Substitution from (6) in the second term leads to

$$\frac{d^2 \mathbf{m}}{dt'^2} = \frac{2\mu}{\hbar} \ \mathbf{m} \times \frac{d\mathbf{B}'(t')}{dt'} + \frac{2\mu}{\hbar} \left( \mathbf{m} \times \mathbf{B}'(t') \right) \times \mathbf{B}'(t') \tag{8}$$

If the magnetic field were constant, only the second term in (8) would be present. As pointed out privately by Adrian Melissinos, this term yields the classical equivalent of the spin-flip radiation discussed in the introduction of ref. 3. But here we are interested in the opposite

---

[3] J. D. Jackson, "On understanding spin-flip synchrotron radiation and the transverse polarization of electrons in storage rings," Rev. Mod. Phys. **48**, 417-433 (1976). See Eqs. (7) -(10) and remarks in section B on p. 420.

---

limit, that of a rapidly varying magnetic field that produces only a very small oscillating term in the magnetic moment. Dropping the second term in (8, approximating **m** on the right hand side with the static part **m**(0), and writing $\mathbf{B'}(t') = \hat{\mathbf{x}} B'(t')$, the adequately approximate solution for the second time derivative of **m** is

$$\frac{d^2\mathbf{m}}{dt'^2} = \frac{2\mu}{\hbar} \left(\mathbf{m}(0) \times \hat{\mathbf{x}}\right) \frac{d B'(t')}{dt'} \tag{9}$$

Then the radiated power per unit solid angle is

$$\frac{d P(t')}{d\Omega'} = \frac{\mu^2}{\pi \hbar^2 c^3} \left|\hat{\mathbf{k}} \times \left(\mathbf{m}(0) \times \hat{\mathbf{x}}\right)\right|^2 \left|\frac{d B'(t')}{dt'}\right|^2 \tag{10}$$

Following the discussion of p. 673-674 of Ref. 2, the radiated intensity per unit frequency interval per unit solid angle in K' is

$$\frac{d^2 I}{d\Omega' d\omega'} = \frac{2}{\pi} \frac{\mu^2}{\hbar^2 c^3} \left|\hat{\mathbf{k}} \times \left(\mathbf{m}(0) \times \hat{\mathbf{x}}\right)\right|^2 \omega'^2 |B'(\omega')|^2 \tag{11}$$

where $B'(\omega')$ is defined by (2) and only positive frequencies are relevant in (11).

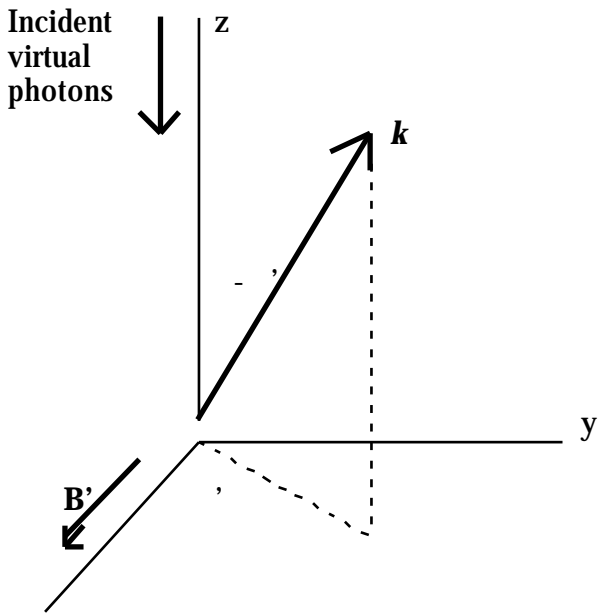

Figure 1 shows the variables for evaluation of the angular distribution. Note that the angle ' is the angle of the radiation with respect to the *negative* z-axis, representing the scattering angle for the outgoing photon, if we view the process as scattering of the virtual quanta.

Figure 1

x

The angular distribution factor in (11) is

$$|\hat{\mathbf{k}} \times (\mathbf{m}(0) \times \hat{\mathbf{x}})|^2 = m_z^2(0)[\cos^2\theta' + \sin^2\theta'\cos^2\phi'] + m_y^2(0)\sin^2\theta'$$

$$- 2m_z(0)m_y(0)\sin\theta'\cos\theta'\sin\phi' \quad (12)$$

The average over azimuthal angle $\phi'$ is

$$\langle |\hat{\mathbf{k}} \times (\mathbf{m}(0) \times \hat{\mathbf{x}})|^2 \rangle_{\phi'} = \frac{1}{2} m_z^2(0)[1 + \cos^2\theta'] + m_y^2(0)\sin^2\theta' \quad (13)$$

Summing over final spin states, we have $m_z^2(0) = m_y^2(0) = \mu^2$. These equalities can be understood in terms of simple spin considerations: (a) $|\mathbf{m}|^2 = |\mu\sigma_x|^2 + |\mu\sigma_y|^2 + |\mu\sigma_y|^2 = 3\mu^2$; (b) the radiation intensity involves the square of the matrix elements, summed over final states, for example, $|m_y(0)|^2 = \sum_f |\langle f|\mu\sigma_y|i\rangle|^2 = \mu^2 \langle i|\sigma_y^2|i\rangle = \mu^2$. The net result for the azimuthally averaged energy radiated per unit frequency per unit solid angle is

$$\frac{d^2 I}{d\Omega' d\omega'} = \frac{1}{\pi} \frac{\mu^4}{\hbar^2 c^3} [3 - \cos^2\theta'] \omega'^2 |B'(\omega')|^2 \quad (14)$$

This is the classical intensity of radiation scattered by the neutrino. Since $c|B'(\omega')|^2/2\pi$ can be viewed as the flux of energy of the virtual photons, the classical "scattering" cross section is

$$\left.\frac{d\sigma}{d\Omega'}\right|_{classical} = \frac{2\mu^4 \omega'^2}{\hbar^2 c^4} [3 - \cos^2\theta'] \quad (15)$$

**C. Compton-recoil modification**

When the energy of the incident photon becomes comparable to the rest energy of the particle, the particle's recoil makes the energy (frequency) of the scattered photon less than that of the incident photon. That's the effect discovered by Compton. Just as Jackson (Ref. 2) does in Section 14.8 for the scattering of radiation by a charged particle (to convert the Thomson cross section into the quantum-mechanical result for scalar particles), we take account of the recoil kinematics by multiplying the classical expressions (14) or (15) by the "Compton factor,"

$$\left(\frac{\omega'_{scatt}}{\omega'_{inc}}\right)^2 = \left(1 + \frac{\hbar\omega'}{m_\nu c^2}(1 - \cos\theta')\right)^{-2} \qquad (16)$$

Eq.(16) is just the ratio of the emitted photon phase space, with and without recoil correction.

As described in the Introduction, the typical virtual photon has an energy large compared to the neutrino's rest energy. The Compton factor vanishes rapidly as the angle œ' increases. In the approximation that the spectrum is dominated by frequencies such that $\hbar\omega' >> m_\nu c^2$, we can put $\cos^2\theta' = 1$ in the numerator. Then the integration of (14) over angles becomes

$$\frac{dI}{d\omega'} = \frac{4\mu^4 \omega'^2}{\hbar^2 c^3}|B'(\omega')|^2 \frac{m_\nu c^2}{\hbar\omega'} \int_0^\infty dx/(1+x)^2 = \frac{4\mu^4 m_\nu \omega'}{\hbar^3 c}|B'(\omega')|^2 \qquad (17)$$

where $x = \hbar\omega'(1 - \cos\theta')/m_\nu c^2$ and we have set the large upper limit to infinity (to the neglect of relative terms of order $m_\nu c^2/\hbar\omega'$). The corresponding cross section for incident frequency $\omega'$ is

$$\sigma_{semi-classical} = \frac{8\pi\mu^4 m_\nu \omega'}{\hbar^3 c^2} = \frac{4\pi\mu^4}{\hbar^4 c^4}(s - m_\nu^2 c^4) = \frac{\pi}{4}\alpha^2 \left(\frac{\mu}{\mu_0}\right)^4 \frac{(s - m_\nu^2)}{m_e^4} \qquad (18)$$

In the final form we have reverted to "particle physics" units where $\hbar$ and $c$ are suppressed. The unit of magnetic moment is $\mu_0 = e\hbar/2m_e c$, $2m_\nu c^2 \hbar\omega' = s - m_\nu^2 c^4$, and $s$ is the usual Mandelstam variable.

### D. Number of photons emitted

The distribution in incident frequency $\omega'$ of the number of photons radiated (or more accurately, removed from the virtual photon flux incident) is (17) divided by the energy $\hbar\omega'$ of a single photon;

$$\frac{dN}{d\omega'} = \frac{4\mu^4 m_\nu}{\hbar^4 c}|B'(\omega')|^2 \qquad (19)$$

The total number $N$ of photons, the integral over $\omega'$, is according to (4),

$$N = \frac{2\mu^4}{\hbar^4 c^4} E_\nu B_0^2 L = \frac{\alpha^2}{8}\left(\frac{\mu}{\mu_0}\right)^4 \frac{\hbar^2 c^2}{(m_e c^2)^4} E_\nu B_0^2 L \qquad (20)$$

Here $E_\nu = \gamma m_\nu c^2$ is the neutrino energy in the laboratory and $B_0^2 L$ is the longitudinal integral of the square of the magnetic field given by (1).

**E. Comparison with QED calculation of "Compton" scattering by neutrino**

The lowest order QED cross section, akin to the Klein-Nishina formula, for the scattering of photons from the massive neutrino with a magnetic moment has been evaluated by N. V. Mikheev [Ref. 4]. In invariant form, the differential scattering cross section is

---

[4] N. V. Mikheev, private communication. The total cross section in the approximation of the neglect of the neutrino mass is given as Eq.(8) of Ref. 1.

---

$$\frac{d\sigma}{dt} = 4\pi\mu^4 \left[ \frac{m_\nu^2 - u}{s - m_\nu^2} - \frac{2m_\nu^2 t}{(s - m_\nu^2)^2} - \frac{2m_\nu^4 t^2}{(m_\nu^2 - u)(s - m_\nu^2)^3} \right] \quad (21)$$

Here we have again suppressed powers of $\hbar$ and $c$. For comparison with the classical cross section (15) or the semi-classical product of (15) and (16), we go to the neutrino rest frame and convert to the conventional differential cross section per unit solid angle. The result is

$$\left. \frac{d\sigma}{d\Omega'} \right|_{QED} = \frac{2\mu^4 \omega'^2}{\hbar^2 c^4} \left( 3 - \cos^2\theta' \right) \left( \frac{\omega'_{scatt}}{\omega'_{inc}} \right)^3 \quad (22)$$

or

$$\left. \frac{d\sigma}{d\Omega'} \right|_{QED} = \frac{2\mu^4 \omega'^2 (3 - \cos^2\theta')}{\hbar^2 c^4 \left[ 1 + \hbar\omega'(1 - \cos\theta')/m_\nu c^2 \right]^3} \quad (23)$$

Two features: (I) The low frequency limit agrees completely with the classical result, as expected; (ii) The quantum modification involves the *third* power of $(\omega'_{scatt}/\omega'_{inc})$, not the second, as given by (16). The additional power of the "Compton factor" causes a more rapid decrease with increasing angle and so the integration over angles corresponding to (17) yields a total cross section at high energies one half as large, namely,

$$\sigma_{QED} = \frac{\pi}{8} \alpha^2 \left( \frac{\mu}{\mu_0} \right)^4 \frac{(s - m_\nu^2)}{m_e^4} \quad (24)$$

One might wonder if a more careful consideration of the kinematics could have fixed the modification of Section **C** to give the cube rather than the square of the ratio of outgoing to incoming frequencies and so yield a very satisfying semi-classical derivation of the process. I think not, at least not without prescient "intuition.". The quantum correction for magnetic moment radiation is different from that for charge radiation. The Klein-Nishina cross section gives an illustration for *g = 2*. The differential cross section in the target rest frame can be written

$$\frac{d\sigma}{d\Omega'}\bigg|_{K-N} = \frac{\alpha^2}{m_e^2}\left(\frac{\omega'_{scatt}}{\omega'_{inc}}\right)^2 \left[\frac{1}{2}(1+cos^2\theta') + \frac{\omega'_{scatt}}{\omega'_{inc}}\frac{\omega'^2_{inc}}{2m^2}(1-cos\theta')^2\right] \qquad (25)$$

**The factors in front times the first term is the contribution of the charge, namely, the Thomson cross section times (16). The second term, with its additional power of $\omega'_{scatt}/\omega'_{inc}$ and square of the incident frequency, is the Dirac magnetic moment contribution. The particular angular dependence is characteristic of *g = 2*. The angular dependence in (22) or (23) is characteristic of *g = ∞*.**
     **The semi-classical calculation gives semi-quantitative results, differing by a factor of two from the QED results for $\sigma$ and the number of photons *N*. Since the whole process is of only academic interest at best, the semi-classical physicist aka experimenter can be content that he/she now "understands" the mechanism.**